\documentclass[conference]{IEEEtran}
\usepackage{cite}
\usepackage[pdftex]{graphicx}
\graphicspath{/images/}
\usepackage[cmex10]{amsmath}
\usepackage{algorithmic}
\usepackage{array}
\usepackage{url}

\begin{document}
\title{Code-Expanded Random Access for Machine-Type Communications}

\author{\IEEEauthorblockN{Nuno K. Pratas, Henning Thomsen, \v Cedomir Stefanovi\' c, Petar Popovski}
\IEEEauthorblockA{Department of Electronic Systems, Aalborg University, Denmark \\
Email: nup@es.aau.dk, ht1703@gmail.com and \{cs,petarp\}@es.aau.dk}}

\maketitle

\begin{abstract}
The random access methods used for support of machine-type communications (MTC) in current cellular standards are derivatives of traditional framed slotted ALOHA and therefore do not support high user loads efficiently.
Motivated by the random access method employed in LTE, we propose a novel approach that is able to sustain a wide random access load range, while preserving the physical layer unchanged and incurring minor changes in the medium access control layer.
The proposed scheme increases the amount of available contention resources, without resorting to the increase of system resources, such as contention sub-frames and preambles.
This increase is accomplished by expanding the contention space to the code domain, through the creation of random access codewords.
Specifically, in the proposed scheme, users perform random access by transmitting one or none of the available LTE orthogonal preambles in multiple random access sub-frames, thus creating access codewords that are used for contention.
In this way, for the same number of random access sub-frames and orthogonal preambles, the amount of available contention resources  is drastically increased, enabling the support of an increased number of MTC users.
We present the framework and analysis of the proposed code-expanded random access method and show that our approach supports load regions that are beyond the reach of current systems.
\end{abstract}

\IEEEpeerreviewmaketitle

\section{Introduction}
\label{sec:Introduction}

In the past few years there has been an increase in the number of networked machines on current networks designed for human-centric communications. This has led to a shift in the conventional perception on communications towards device-centric communications, which is independent of human interaction~\cite{5741142,AnthonyLo2011}. While the requirements of traditional human-centric communications are high bit-rates and lower latency, in device-centric communications, the main requirement is the massive transmission of simultaneous low data rate messages.


The main challenge in Machine-to-Machine (M2M) communications, denoted as Machine-Type Communications (MTC) by 3GPP, is to adapt cellular networks to efficiently handle MTC traffic characteristics, specifically the load generated by massive simultaneous low data rate transmissions. 
Within 3GPP, there is an ongoing study on the adaptation of the 3GPP cellular networks to handle the MTC traffic~\cite{TR37.8682011,6211484}, through the inclusion of enhanced load control mechanisms in the Radio Access Network (RAN).
This is paramount, because due to the expected large number of deployed MTC devices, the cellular networks are expected to withstand traffic bursts~\cite{M.ZubairShafiq2012}.
In such situations, radio and signalling network congestions may occur due to mass concurrent transmissions~\cite{3GPPTS22.368}, which can lead to large delays, packet loss and, in the extreme case, service unavailability.

In 3GPP there have been proposed several solutions for managing the random access load~\cite{TR37.8682011}.
These solutions include access class barring~\cite{6093905,6162473}, and back-off schemes, where the network controls the load by restricting and delaying the random access, thus spreading the random access in time and reducing the effect of batch arrivals. Other solutions divide and adapt the amount of random access resources for human-centric and device-centric traffic, such as orthogonal random access~\cite{6162474} and dynamic random access~\cite{AnthonyLo2011}, where overload due device-centric traffic will not affect the human-centric traffic and vice versa.


In this paper we propose a random access method inspired by the LTE random access, which is in line with the dynamic Random Access Channel (RACH) resource allocation approach.
Here the dynamic resource allocation is not accomplished through traditional means, such as the increase of the number of available preambles and/or random access sub-frames, but instead by introducing the concept of code-expanded random access.
The motivation behind this proposal is to enable existing systems to sustain a large and bursty random access loads, while preserving the physical layer unchanged and incurring minimal changes in the medium access control layer.

\begin{figure}
	\centering
		\includegraphics[width=\columnwidth]{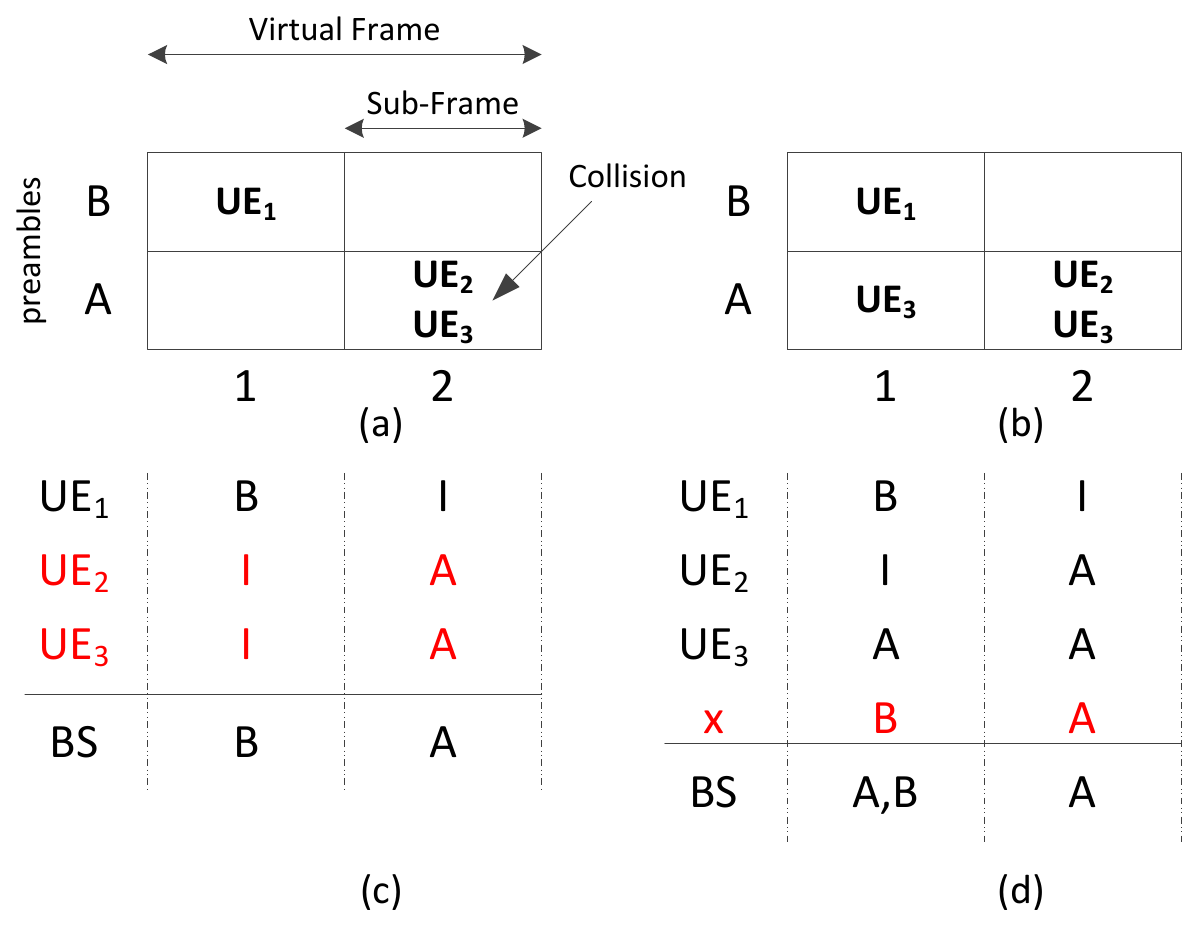}
	\caption{(a) Reference random access, (b) Code-expanded random access, (c) Reference random access codewords, with collisions in red, (d) Code-expanded codewords, with phantom codeword in red.}
	\label{fig:CodeWordExample}
\end{figure}

To illustrate what is meant by code-expanded random access, consider the diagrams in Fig.~\ref{fig:CodeWordExample}, where the selection of preambles and sub-frames in the reference and in the proposed code-expanded random access schemes are depicted.
In the reference scheme, each MTC user (M-UE) performs the random access by selecting one of the available preambles, in this example denoted as \emph{A} and \emph{B}, and then selecting a random access sub-frame to transmit the chosen preamble, as depicted in Fig.~\ref{fig:CodeWordExample}(a).
The Base Station (BS) then discerns between M-UEs according to the observed preambles in each random access sub-frame, as shown in Fig.~\ref{fig:CodeWordExample}(c).
In the proposed scheme, depicted in Fig.~\ref{fig:CodeWordExample}(b), the random access sub-frames are grouped in \emph{virtual frames} and the random access is performed on the virtual frame level.
In each sub-frame of the virtual frame, the M-UE transmits one or none of the available preambles (we assume that in the case when none of the preambles is transmitted, the M-UE actually transmits an \emph{idle} preamble, henceforth denoted as \emph{I}), thus creating an \emph{access codeword} with length equal to the length of the virtual frame.
In this case, the BS observes the distribution of preambles over the virtual frame sub-frames, i.e., observes a set of access codewords that are then used to discern between M-UEs.

The proposed code-expanded random access scheme is a generalization of the reference scheme.
In the reference scheme the codewords are composed of just one preamble sent in one of the sub-frames of the virtual frame, with the remaining sub-frames at idle.
In the proposed scheme, each M-UE selects a codeword consisting of a randomly chosen preamble (including the idle preamble) per every sub-frame  of the virtual frame.
In this way, the number of contention resources is expanded and the amount of collisions is reduced - a collision occurs when two or more M-UEs select the same codeword, as depicted in Fig.~\ref{fig:CodeWordExample}(c).

On the other hand, while reducing the amount of collisions, the code-expanded generalization introduces a shortcoming not present in the reference scheme.
In the reference scheme, the codewords do not introduce ambiguities at the BS in regards to which codewords where transmitted, i.e., based on the observation of the preambles in the virtual frame, the BS can always discern which codewords were actually sent.
In the proposed scheme, such ambiguities exist - based on the observation, the BS may conclude that there are more codewords present in the virtual frame than there were actually sent.
Namely, multiple combinations of transmitted codewords can yield the same observation, introducing phantom codewords which were not sent by any of the transmitting M-UEs.
In Fig.~\ref{fig:CodeWordExample}(d) is depicted a combination of the three codewords used by M-UEs, which misleads the BS to perceive the phantom codeword $(B,A)$. 
The existence of phantom codewords affects adversely the performance, however, we show that this is significant only when the network is experiencing low user loads.
Specifically, we show that the efficiency of the proposed approach (i.e., the fraction of M-UEs that successfully contended for access) for moderate and high loads and for the same number of used preambles and sub-frames in the virtual frame, is substantially higher than the efficiency of the reference scheme, despite the drawbacks caused by phantom codewords.
We also show that by choosing the operating random access method according to the user load, it is possible to maintain an efficient random access over a large load region using the same number of preambles and sub-frames.

The remainder of this paper is organized as follows.
In Section~\ref{sec:SystemModel} we model and analyse both the reference and the proposed random access scheme.
In Section~\ref{sec:AdaptiveCodeExpandedRandomAccess} we discuss how the proposed code-expanded scheme can be used to enable a random access mechanism which adapts according to the load in the random access.
Finally, Section~\ref{sec:Conclusion} concludes the paper.

\section{System Model}
\label{sec:SystemModel}


\subsection{Preliminaries}
\label{sec:Preliminaries}

In LTE the random access is performed through the Random Access Channel (RACH), which is mapped in the physical layer to the Physical Random Access Channel (PRACH)~\cite{3GPPTS36.212}.
To start the asynchronous access each M-UE first selects one of the designated preambles and then waits until the next available PRACH sub-frame.
The M-UE then performs the random access procedure, as described in~\cite{3GPPTS36.321,3GPPTS36.213}.
The periodicity of the PRACH resources can scale according to the expected RACH load. Therefore, the PRACH resources can occur from once in every sub-frame to once in every two frames, where a frame consists of 10 sequential sub-frames~\cite{3GPPTS36.211}.

\begin{figure}
	\centering
		\includegraphics[width=\columnwidth]{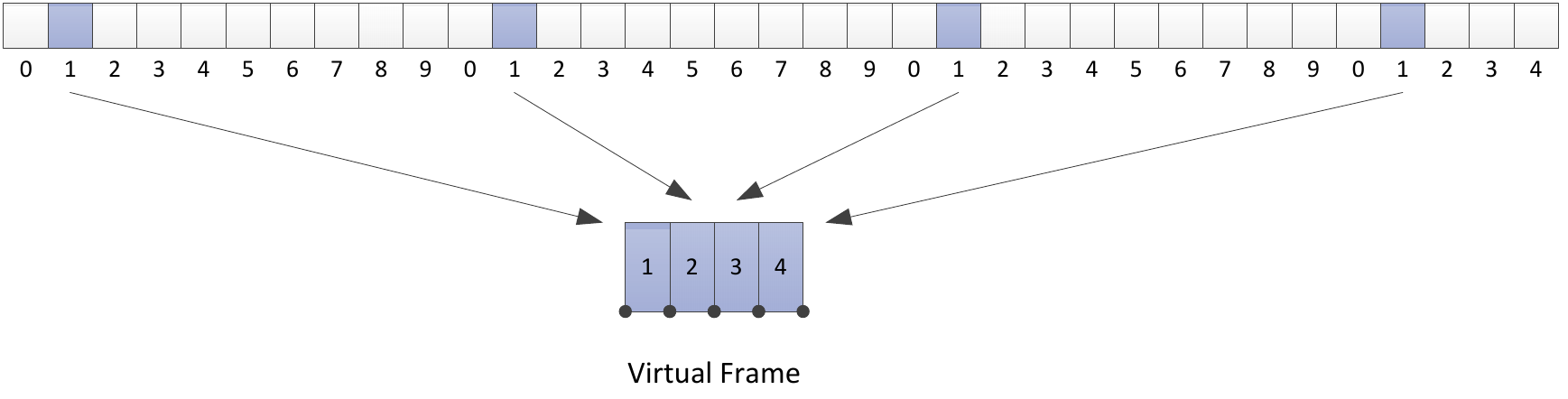}
	\caption{Example of PRACH opportunities organized in virtual frames.}
	\label{fig:LTE_RandomAccess_VirtualFrames}
\end{figure}

As outlined in Section~\ref{sec:Introduction}, we assume that a group of consecutive sub-frames is organized in a virtual frame reserved for PRACH (i.e., random access), as depicted in the Fig.~\ref{fig:LTE_RandomAccess_VirtualFrames}. 
We denote the number of sub-frames within a virtual frame by $L$ and the number of available preambles by $M$.

\subsection{Reference Random Access Scheme}
\label{sec:ReferenceRandomAccess}

\begin{figure}
	\centering
		\includegraphics[width=\columnwidth]{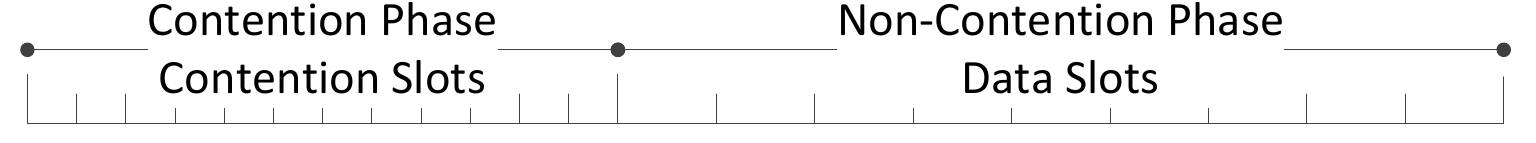}
	\caption{Access reservation.}
	\label{fig:AccessReservation}
\end{figure}

The reference scheme considered here is inspired by the LTE random access ~\cite{3GPPTS36.321,3GPPTS36.213}. 
We model it as an access reservation scheme~\cite{Bertsekas1987}, as depicted in Figure~\ref{fig:AccessReservation}.
We focus on the contention phase, which consists of $L$ sub-frames (contention slots), and these sub-frames constitute a single virtual frame.
We assume that there are $N$ contending M-UEs; each M-UE randomly choses a preamble and a sub-frame of the virtual frame in which the chosen preamble is sent.\footnote{In the LTE scenario, there is no notion of virtual frames and a M-UE can send a preamble in every sub-frame, in contrast to the reference scenario presented here, where a preamble is sent once per virtual frame. However, one could argue that, for high user loads, the efficiency of the contention phase of the LTE scenario is worse than of the reference scheme, as the number of collisions in the former is higher.}

The fraction of the M-UEs that succeed in the contention phase is denoted as the efficiency of the contention phase $S_{r}(N)$, and it is given by:
\begin{equation}
 S_{r}(N) = \frac{N_S}{N_S + N_C},
 \label{Sr}
\end{equation}
where $N_S$ is the number of codewords used by a single M-UE (singles) and $N_C$ is the number of codewords used by multiple M-UEs (collisions).
From now on, we assess $N_S$ and $N_C$ by their expected values.

Assuming that the number of M-UEs contending per slot is modeled by the random variable $X$, then the probability that in a given contention slot there are $k$ M-UEs contending is:
\begin{equation}
	\Pr[X = k] = \binom{N}{k} \left(\frac{1}{A_r}\right)^k \left(1-\frac{1}{A_r}\right)^{N-k},
	\label{SlotDistribution}
\end{equation}
where $A_r$ is the number of codewords available in the reference scheme.
As outlined earlier, in a contention phase consisting of $L$ sub-frames, each M-UE sends only one preamble, therefore the number of available codewords is given by:
\begin{equation}
	A_{r} = M \cdot L.
	\label{Ar}
\end{equation}
The expected number of codewords chosen by a single M-UE, $N_S$, is given by:
\begin{align}
	N_S = \Pr[X = 1] \cdot A_{r} = N \left(1-\frac{1}{A_{r}}\right)^{N-1},
										 \label{NSr}
\end{align}
and, similarly, the expected number of codewords chosen by multiple M-UEs, $N_C$, is:
\begin{align}
	N_C & = \Pr[X > 1] \cdot A_{r} = \left[ 1 - \Pr[X = 0] - \Pr[X = 1] \right] \cdot A_{r} \nonumber \\
										 & = \left[ 1 - \left(1-\frac{1}{A_{r}}\right)^{N} - \frac{N}{A_{r}} \left(1-\frac{1}{A_{r}}\right)^{N-1}\right] A_{r}.
	\label{NCr}
\end{align}

Substituting~(\ref{NSr}) and~(\ref{NCr}) in~(\ref{Sr}), we get the estimate of the efficiency of the contention phase.

\subsection{Code-Expanded Random Access}
\label{sec:CodeExpandedRandomAccess}



In the code-expanded random access, the outcome of the contention phase is the set of the codewords perceived by the BS, and all the perceived codewords are assumed to belong to the M-UEs that successfully contended for the access.
However, the set of perceived codewords actually contains the codewords that are used just by a single M-UE (singles), codewords that are used by multiple M-UEs (collisions) and codewords that are used by none of the M-UEs (phantom codewords, see Fig.~\ref{fig:CodeWordExample}).  
The efficiency of the code-expanded random access $S_{e}(N)$ is:
\begin{equation}\label{Se}
	S_{e}(N) = \frac{N_S}{N_P}.
\end{equation}
where the $N_S$ is the number of single codewords, and $N_P$ is the expected number of codewords that the BS perceives, i.e., $N_P$ accounts for all single, collision and phantom codewords, where $N_P\leq A_{e}$.

The distribution of devices contending per virtual frame is modeled in the same way as in~(\ref{SlotDistribution}), with the difference that the number of available codewords is now higher.
In the code-expanded approach, the M-UEs send in each sub-frame of the virtual frame either one of the $M$ preambles or the idle preamble $I$, so the total number of available codewords is:
\begin{equation}
	A_e = \left(M + 1\right)^L-1,
	\label{Ae}
\end{equation}
where the all-idle codeword is excluded.

The expected number of codewords chosen by a single M-UE, $N_S$, is given by:
\begin{align}
	N_S & = \Pr[X = 1] \cdot A_{e} = N \left(1-\frac{1}{A_{e}}\right)^{N-1}.
	\label{NSe}
\end{align}
The method to obtain $N_P$ is discussed in the next subsection.

\subsection{Calculation of $N_P$}
\label{sec:BetaNAsymptoticModelling}


For the calculation of $N_P$ we use a representation based on a Markov Chain (MC) that describes the evolution of the perceived number of codewords by the BS when the number of contending M-UEs increases sequentially from 1 to $N$.
We note that it is assumed that the M-UEs select their codewords independently and uniformly at random from the set of available codewords.
The MC states are determined by the configuration that corresponds to the number of the observed preambles in the sub-frames of the virtual frame, which is created by the actual codewords selected by the M-UEs.
The configuration is denoted by $(C_1, C_2, ..., C_L)$, where $C_j$ is the number of observed preambles by the BS in the $j$-th sub-frame.
We note here that $C_j$ always includes the idle preamble, i.e., if the number of unique preambles in the $j$-th sub-frame is $C_j$, then the number of actually observed preambles is $C_j - 1$.
Each state is characterized by a cardinality, which is the cardinality of the set of codewords perceived by the BS that is created by the given configuration.
For $i$-th state configuration $(C_1, C_2, C_3, ..., C_L)$ the corresponding cardinality is:
\begin{equation}
	\alpha_i=\prod_{j=1}^L C_j - 1,
	\label{alpha}
\end{equation}
where, once again, we assumed that all-idle codeword is not used by any of the M-UEs.
Using this model, $N_P$ can be assessed as the average cardinality of the set of the codewords perceived after $N-1$ transitions of MC, averaged over the probability distribution of the MC states after $N-1$ transitions.

%
\begin{table}[t]
	\centering
			\begin{tabular}{| c || c | c |}
		  	\hline                        
  			Codeword& \multicolumn{2}{c|}{L} \\
  					 & 1 & 2 \\
  			\hline
  			1 & I & A \\
  			2 & I & B \\
  			3 & A & I \\
  			4 & A & A \\
  			5 & A & B \\
  			6 & B & I \\
  			7 & B & A \\
  			8 & B & B \\
  			\hline  
			\end{tabular}
	\caption{Codebook, $L=2$, $M=2$.}
	\label{tab:CodebookL2OS2}
\end{table}
To ease the explanation, we focus on an example case where $L=2$ and $M=2$, and therefore $A_{e} = 8$.
The full codebook is shown in Table~\ref{tab:CodebookL2OS2}, while the MC representation including the state configurations, cardinalities and possible state transitions is shown in Table~\ref{tab:FullCodebookCardinalityStateTransitionsL2OS2}. For example, the state configuration of the state $2$ is $(1,3)$, implying that in the first sub-frame of the virtual frame there is an idle preamble $I$, and in the second sub-frame there is an idle preamble and both available preambles, \emph{A} and \emph{B}.
The cardinality of state 2 is then $\alpha_2 = 2$.

Initially, when there is only one M-UE attempting random access, the system can be in the states $1$, $3$ and $4$.
The probability of the system being in any of those states is simply the ratio of the number of codewords from the codebook which provide the corresponding state configuration and the total number of available codewords $A_e$.
For example, for state $1$, where the state configuration is $(1,2)$, from Table~\ref{tab:CodebookL2OS2} it can be seen that there are only two codewords that satisfy the state configuration, which are $(I,A)$ and $(I,B)$. Therefore, the probability of the BS perceiving this state upon the M-UE transmission of one of the available codewords is $\frac{2}{8}=\frac{1}{4}$.
A similar reasoning is done for the remaining possible initial states, and the following initial state probability vector $\mathbf{\pi}^{(1)}$ is:
\begin{equation}
	\mathbf{\pi}^{(1)} = \begin{bmatrix}
		1/4 & 0 &	1/4 & 2/4 & 0 &	0 &	0 &	0 \\		
		\end{bmatrix}
\end{equation}
where $\pi^{(1)}_{i}$ is the probability that the system is initially in $i$-th state.
Therefore, when one M-UE attempts random access, i.e. when $N = 1$, $N_P$ is obtained by:
\begin{equation}
	N_P = \sum_{i=1}^{A_{e}} \alpha_i \cdot \pi^{(1)}_i = 2
\end{equation}
where $\alpha_i$ is the cardinality of the $i$-th state, obtained from~(\ref{alpha}) and listed in Table~\ref{tab:FullCodebookCardinalityStateTransitionsL2OS2}.

In case where there are two M-UEs attempting random access, it is assumed that the selection of the transmitted codewords is sequential and independent. Therefore, the codeword selected by the first M-UE leads the system to be in one of the possible initial states, which are $1$, $3$ and $4$. When the second M-UE selects the codeword, the system can transit to any of the states listed in Table~\ref{tab:FullCodebookCardinalityStateTransitionsL2OS2}.
\begin{table}[t]
	\centering
			\begin{tabular}{| c || c | c || c | c |}
		  	\hline                        
  			State IDs & \multicolumn{2}{c||}{Configuration} & Cardinality & Transitions \\
  			  & 1 & 2 & &\\ \hline
  			1 & 1 & 2 & 1 & 1,2,4,5\\
  			2 & 1 & 3 & 2 & 2,5\\
  			3 & 2 & 1 & 1 & 3,4,6,7\\
  			4 & 2 & 2 & 3 & 4,5,7,8\\
  			5 & 2 & 3 & 5 & 5,8\\
  			6 & 3 & 1 & 2 & 6,7\\
  			7 & 3 & 2 & 5 & 7,8\\
  			8 & 3 & 3 & 8 & 8\\
  			\hline  
			\end{tabular}
	\caption{Markov Chain Model, $L=2$, $M=2$.}
	\label{tab:FullCodebookCardinalityStateTransitionsL2OS2}
\end{table}

The transition probabilities can be obtained following the same reasoning as the one used to obtain $\mathbf{\pi}^{(1)}$.
Consider that the first M-UE selects a codeword that puts the system in state $1$, i.e. the M-UE selected either $(I,A)$ or $(I,B)$. Now, whatever codeword the second M-UE selects, the system can only transit to states $1,2,4,5$.
For the system to transit from state $1$ to state $5$, it means that the second M-UE has to select a codeword that consists of either preamble \emph{A} or \emph{B} in the first sub-frame, and the remaining yet unused preamble in the second sub-frame (i.e, the codeword $(A,B)$ or $(B,B)$  if the initial codeword was $(I,A)$, or the codeword $(A,A)$ or $(B,A)$ if the initial codeword was $(I,B)$, thus making the configuration become $(2,3)$).
From the preceding discussion, we see that no matter which codeword caused the chain to be in state $1$, the transition to state $5$ can be caused by the second M-UE selecting (one of) two codewords from the set of all codewords. Therefore, this transition probability is equal to $\frac{2}{8}=\frac{1}{4}$ and it does not depend on which codeword the first M-UE selected to bring the system to state $1$. Using similar reasoning, it can be shown that the transition probabilities do not depend on the choices of codewords that brought the system to a given state and, for every state transition, the transition probability is the ratio of the number of codewords that enable the transition and the total number of available codewords $A_e$.
For the considered example, the transition matrix $\mathbf{P}$ is: 
\begin{equation*}
	\mathbf{P} = \begin{bmatrix}		
				\frac{1}{8} & \frac{1}{8} & 					0 & \frac{4}{8} & \frac{2}{8} & 					0 & 					0 & 					0 \\
 									0 & \frac{2}{8} & 					0 & 					0 & \frac{6}{8} & 					0 & 					0 & 					0 \\		
 									0 & 					0 & \frac{1}{8} & \frac{4}{8}	& 					0 & \frac{1}{8} & \frac{2}{8} & 					0 \\		
 									0 & 					0 & 					0 & \frac{3}{8}	& \frac{2}{8} & 					0 & \frac{2}{8} & \frac{1}{8} \\		
 									0 & 					0 & 					0 & 					0	& \frac{5}{8} & 					0 & 					0 & \frac{3}{8} \\
 									0 & 					0 & 					0 & 					0	& 					0 & \frac{2}{8} & \frac{6}{8} & 					0 \\
 									0 & 					0 & 					0 & 					0	& 					0 & 					0 & \frac{5}{8} & \frac{3}{8} \\							
 									0 & 					0 & 					0 & 					0	& 					0 & 					0 & 					0 & 					1 \\							
	\end{bmatrix}
\end{equation*}
where $P_{ij}$ is the transition probability from $i$-th to $j$-th state.

The expected number of perceived codewords $N_P$ can be obtained as follows:
\begin{equation}
	N_P = \sum_{i=1}^{A_{e}} \alpha_i \cdot \pi^{(N)}_i,
	\label{betaN}
\end{equation}
where $\pi^{(N)}_i$ is the $i$-th element of the state vector $\mathbf{\pi}^{(N)}$, i.e., the probability of the BS perceiving a state configuration given by the $i$-th state after the $N$-th user has chosen his codeword. The $\alpha_i$ is the cardinality of the $i$-th state, given by~(\ref{alpha}).
As for any MC, the state probability distribution $\mathbf{\pi}^{(N)}$ is:
\begin{equation}
	\pi^{(N)} = \pi^{(1)} \cdot \mathbf{P}^{(N-1)}.
\end{equation}

\begin{figure}
	\centering
		\includegraphics[width=\columnwidth]{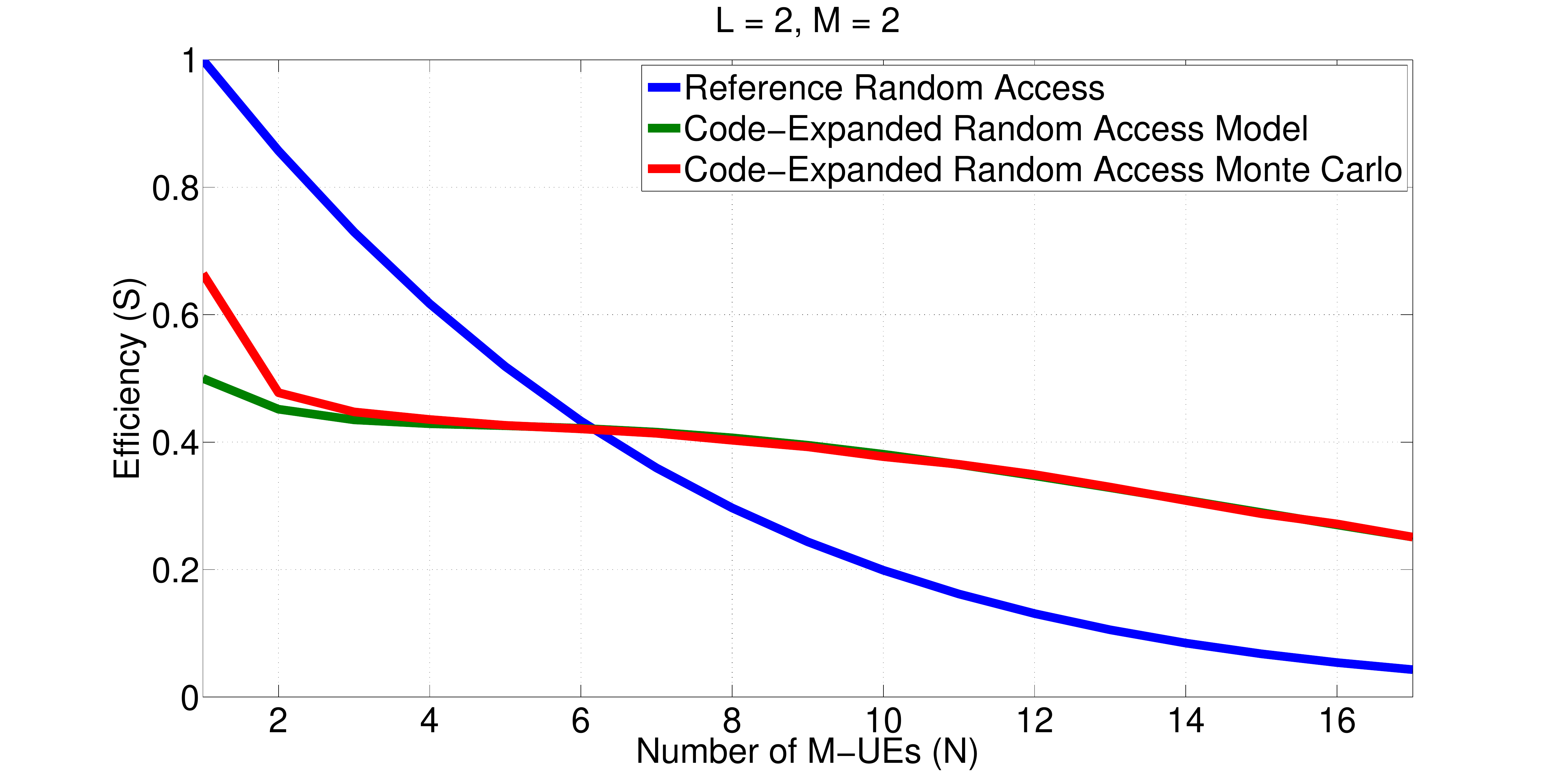}
	\caption{Reference random access and code-expanded random access comparison.}
	\label{fig:EfficiencyWithAsymptoticBetaN}
\end{figure}

Fig.~\ref{fig:EfficiencyWithAsymptoticBetaN} depicts the comparison of the efficiencies for the reference scheme calculated using (\ref{Sr}), code-expanded scheme calculated using~(\ref{Se}) and (\ref{betaN}), and code-expanded scheme that is obtained using Monte Carlo simulations, for the example case when $L=2$ and $M=2$.
It is obvious that, as the number of M-UEs grows, the efficiency of the code-expanded method outperforms the reference one.
Also, both curves corresponding to the code-expanded method converge, justifying the use of (\ref{betaN}) when computing the efficiency.

Finally, we note that the described methodology for modelling $N_P$ can be generalized for arbitrary $L$ and $M$.

\section{Adaptive Code-Expanded Random Access}
\label{sec:AdaptiveCodeExpandedRandomAccess}

In the previous subsection we considered using only the codebook consisting of all the available codewords (i.e., the full codebook).
In this subsection we consider the case when a subset of the full codebook is used, where a subset is obtained by restricting the number of preambles that could be used in each of the sub-frames of the virtual frame (we note that these restrictions are not necessarily the same for every sub-frame).
The efficiency in this case can be computed in the same way as in the previous subsection; we omit the corresponding analysis due to space constraints.

\begin{table}[t]
	\centering
			\begin{tabular}{| c || c | c || c | c |}
		  	\hline                        
  			State & \multicolumn{2}{c||}{Configuration} & Cardinality & Transitions \\
  			   & 1 & 2 & &\\ \hline
  			1  & 1 & 2 & 1 & 1,2,5,6\\
  			2  & 1 & 3 & 2 & 2,3,7\\
  			3  & 1 & 4 & 3 & 3,4,8\\
  			4  & 1 & 5 & 4 & 4,9\\
  			5  & 2 & 1 & 1 & 5,6,10,11\\
  			6  & 2 & 2 & 3 & 6,7,11,12\\
  			7  & 2 & 3 & 5 & 7,8,12,13\\
  			8  & 2 & 4 & 7 & 8,9,13,14\\
  			9  & 2 & 5 & 9 & 9,14\\
  			10 & 3 & 1 & 2 & 10,11,15,16\\
  			11 & 3 & 2 & 5 & 11,12,16,17\\
  			12 & 3 & 3 & 8 & 12,13,17,18\\
  			13 & 3 & 4 & 11 & 13,14,18,19\\
  			14 & 3 & 5 & 14 & 14,19\\
  			15 & 4 & 1 & 3 & 15,16,20,21\\
  			16 & 4 & 2 & 7 & 16,17,21,22\\
  			17 & 4 & 3 & 11 & 17,18,22,23\\
   			18 & 4 & 4 & 15 & 18,19,23,24\\
  			19 & 4 & 5 & 19 & 19,24\\
  			20 & 5 & 1 & 4 & 20,21\\
  			21 & 5 & 2 & 9 & 21,22\\
  			22 & 5 & 3 & 14 & 22,23\\
  			23 & 5 & 4 & 19 & 23,24\\
  			24 & 5 & 5 & 24 & 24\\
  			\hline
			\end{tabular}
	\caption{Markov Chain Model, $L=2$, $M=4$.}
	\label{tab:FullCodebookCardinalityL2OS4}
\end{table}

\begin{figure}[t]
	\centering
		\includegraphics[width=\columnwidth]{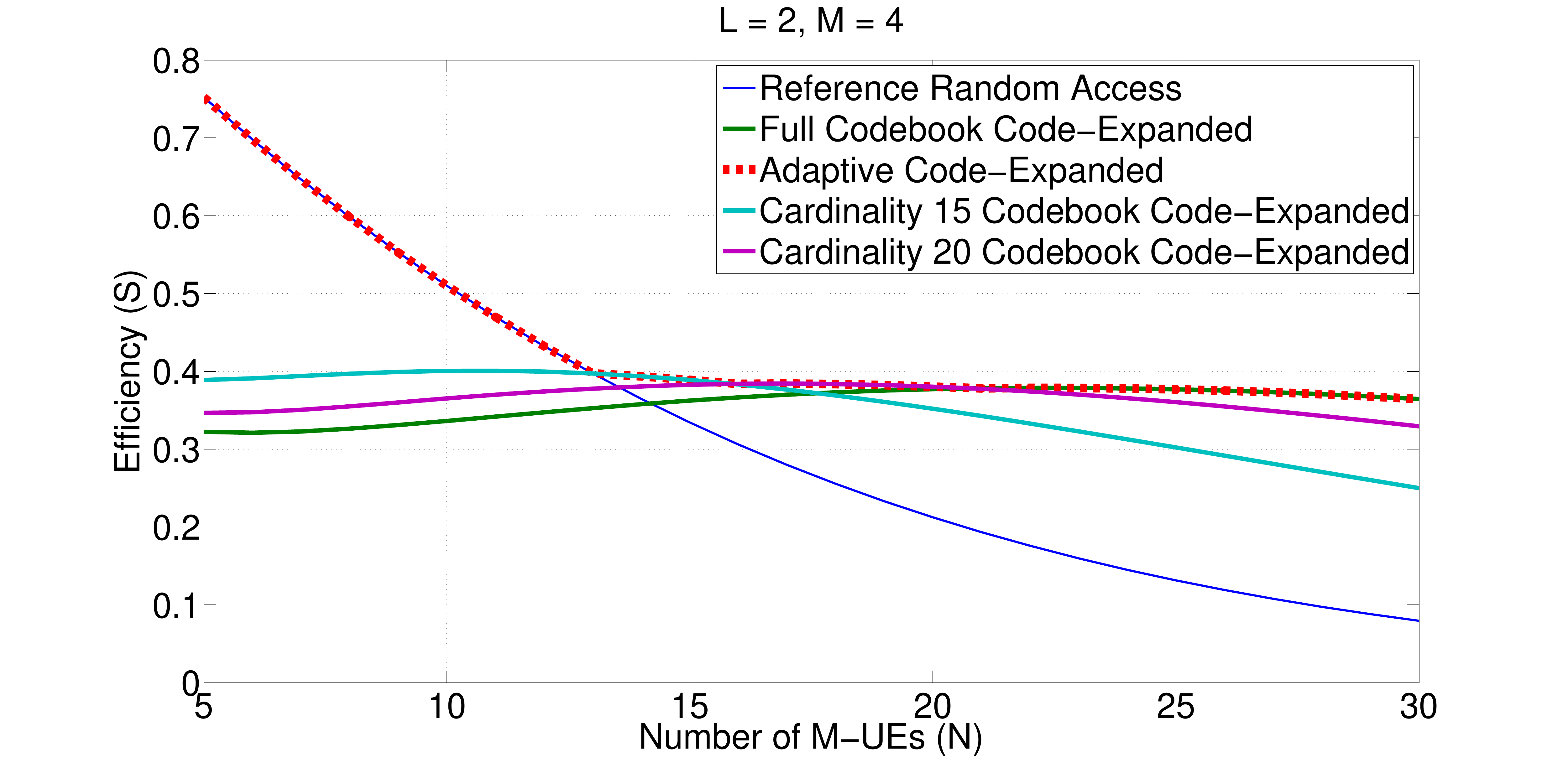}
	\caption{Adaptive Random Access example with $L = 2$ and $M = 4$.}
	\label{fig:ExampleL2Os4}
\end{figure}

As an illustrative example consider the case where $L=2$ and $M=4$, for which the MC model is given in Table~\ref{tab:FullCodebookCardinalityL2OS4}.
The possible cardinality values of the states for this MC are then $\left\{1,2,3,4,5,7,8,9,11,14,15,19,24\right\}$.
This sequence represents all the possible subsets in regards to the number of codes that can be used.
For $L=2$ and $M=4$ the number of codewords for the reference scheme is $A_{r} = 8$, so the cardinalities of interest are the ones where $A_{e} = \left\{9,11,14,15,19,24\right\}$.
In Fig.~\ref{fig:ExampleL2Os4} is depicted the efficiency for the given example, both for the reference scheme and for the code-expanded scheme when cardinalities of interest are used.
As it can be observed, there is a set of threshold values in regards to the number of M-UEs, where the system should increase the cardinality of the codebook in use, in order to maintain the efficiency.
Overall, it could be concluded that an adaptive approach, where the codebook is selected based on the estimate of the user load is the preferred operating strategy.

\begin{figure}[t]
	\centering
		\includegraphics[width=\columnwidth]{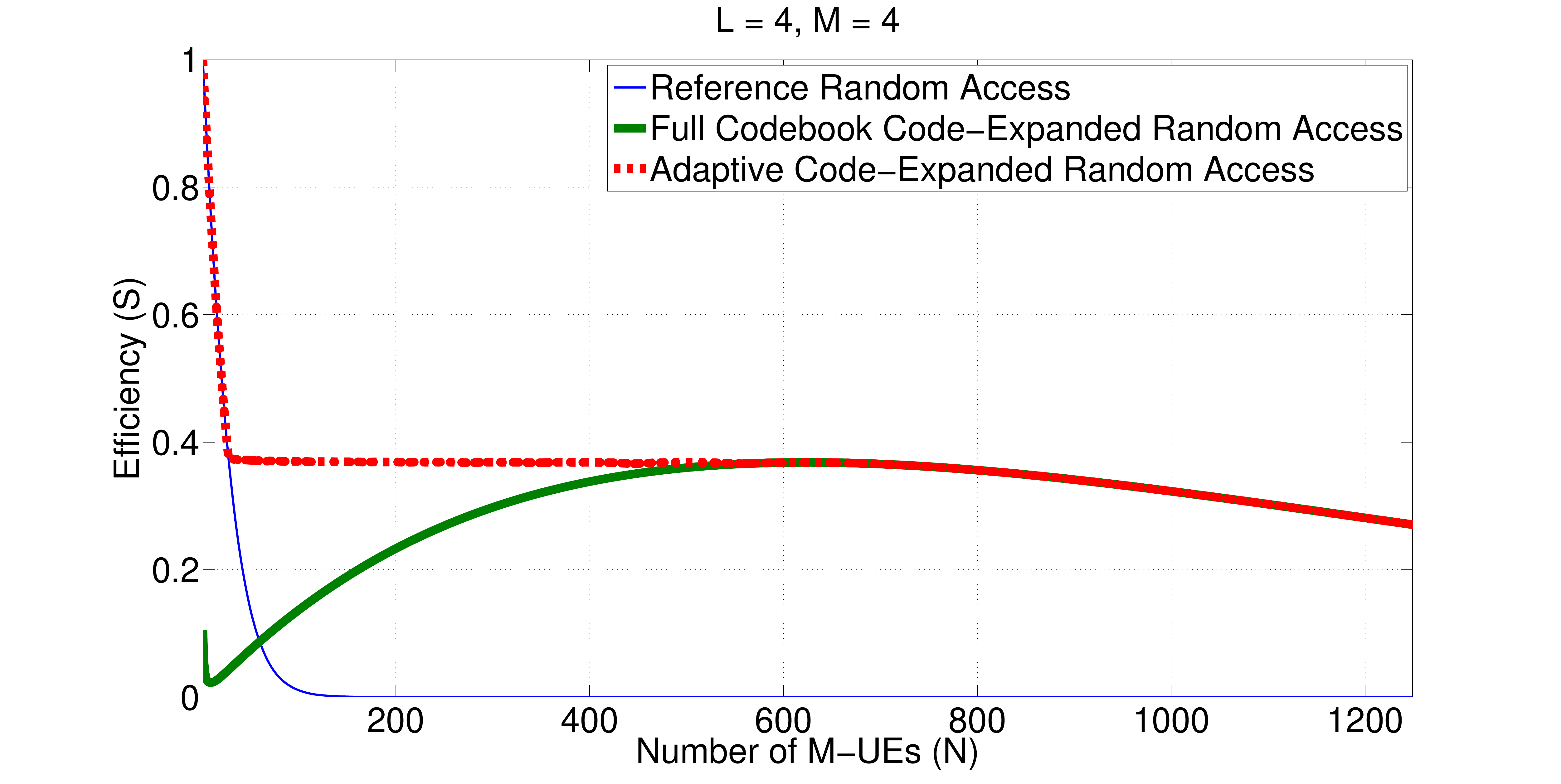}
	\caption{Adaptive Random Access example with $L = 4$ and $M = 4$.}	
	\label{fig:EL4Os4}
\end{figure}

In Fig.~\ref{fig:EL4Os4} is depicted the efficiency of the adaptive code-expanded approach for $L=4$ and $M=4$.
Comparing these results with the results given in Fig.~\ref{fig:ExampleL2Os4}, it can be observed that the region where the gain of the adaptive approach over the case when the full codebook is used is more obvious.
Also, the increased $M$ provides for significant expansion of the supported user load region.

Another way to exploit the code-expanded scheme is to use a significantly lower number of preambles per sub-frame than in the reference scheme, while reaching the same efficiency, since the preamble detection performance decreases with the number of simultaneous preambles in a sub-frame~\cite{R1-0621392006}. 
In the code-expanded scheme the number of available codewords scales exponentially with the virtual frame length~(\ref{Ae}) while in the reference schemes it scales linearly~(\ref{Ar}). Assuming that the number of preambles in use in the reference and code-expanded schemes are denoted as $M_r$ and $M_e$, respectively, the lower bound on $M_e$ required to outperform the reference random access occurs when $A_e > A_r$ and can be calculated using~(\ref{Ae}) and~(\ref{Ar}) as:
\begin{align}
		M_e > \left\lceil \sqrt[L]{M_r \cdot L + 1}-1\right\rceil.
		\label{Me}
\end{align}

\begin{figure}[t]
	\centering
		\includegraphics[width=\columnwidth]{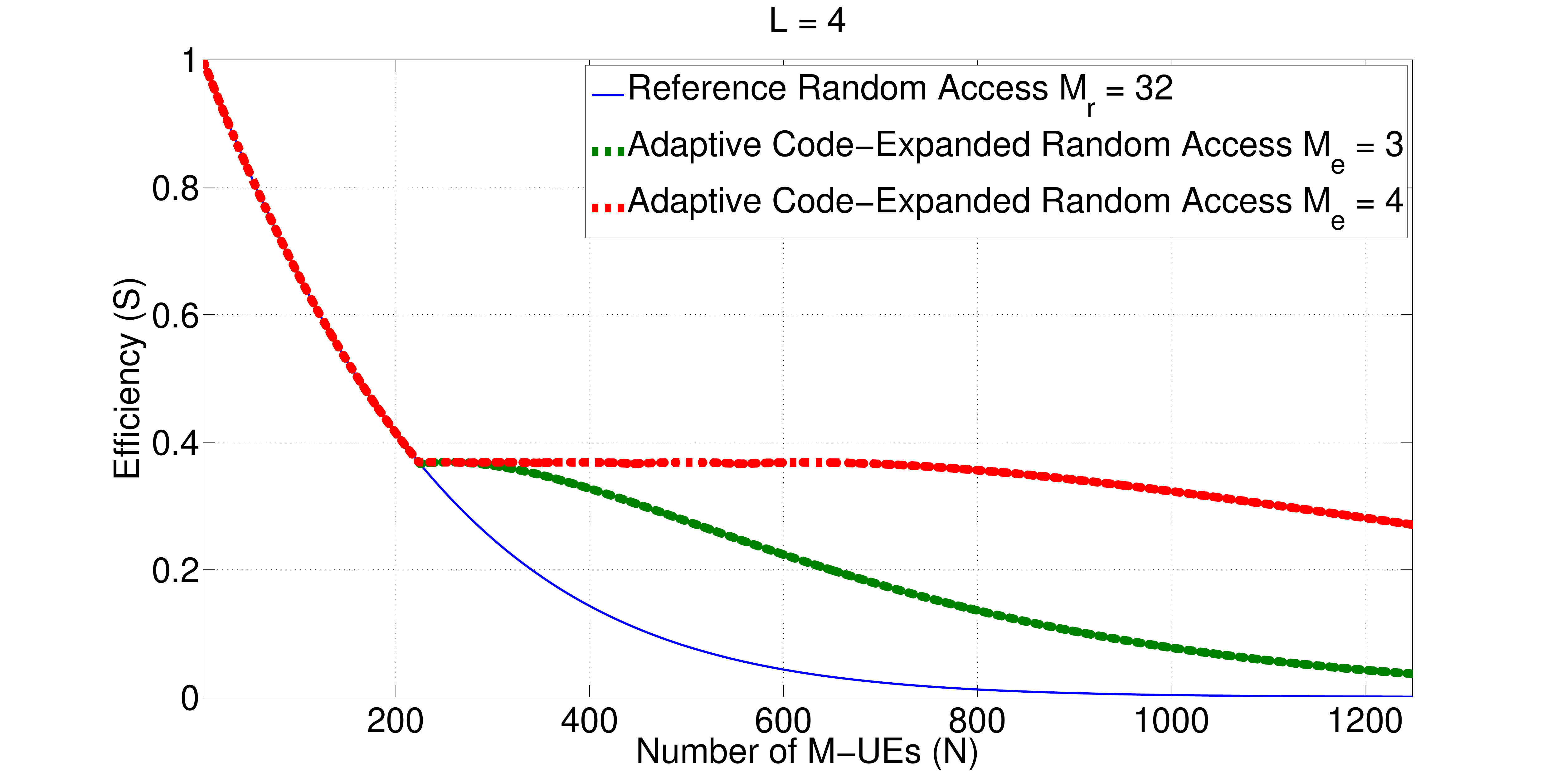}
	\caption{Application Example, $L=4$, $M_r=32$ and $M_e\in{3,4}$.}
	\label{fig:AdaptiveRandomAccessScalabilityL4Os32}
\end{figure}

As example consider the plots in Figures~\ref{fig:AdaptiveRandomAccessScalabilityL4Os32} where $L=4$, $M_r=32$ and $M_e\in\{3,4\}$ ($M_e=3$ is the lower bound obtained by~(\ref{Me})).
In case when $M_e=3$ the code-expanded random access achieves higher performance than the reference scheme when number of users is $N>225$, and the supported user load region is increased several times.
If the number of available preambles is increased just by one, i.e, $M_e=4$, the threshold value for $N$ stays the same (which is in line with results presented in Fig.~\ref{fig:ExampleL2Os4}), but the supported load region is extended significantly further.

The above results suggest the approach in which subsets of available preambles are reserved for different user classes, thus allowing user separation.
The number of preambles per class should be selected according to expected user loads and traffic types, such that a satisfactory efficiency is achieved.

\section{Conclusion}
\label{sec:Conclusion}

In this paper we proposed a code-expanded random access inspired by the LTE random access.
The proposed scheme increases the amount of available contention resources, without resorting to the increase of system resources, such as contention sub-frames and preambles.
This increase is accomplished by expanding the contention space to the code domain, through the creation of random access codewords.

It was shown that for high user loads the proposed scheme is significantly more efficient than the reference scheme.
Also, it was shown that by selecting the appropriate number of random access codewords it is possible to maintain the random access scheme efficiency over a large load region.
This suggests the usage of an adaptive random access scheme, i.e., a combination of the reference and the code-expanded random access, which allows to maintain the efficiency both for low and high user load regions.    

The proposed random access scheme can be further complemented by creating user classes, which are separated by the selection of used preambles. 
Specifically, due to amount of contention codewords that can be obtained with a low number of preambles and that could be reserved for MTC users, the remaining preambles could be used by the human-centric users, thus allowing human-centric and device-centric traffic to coexist.

Finally, we note that for optimal operation of the proposed expansions, it is necessary to obtain the estimates of the user loads.
The design of the adaptive random access scheme that implements the estimation of the user load is a topic of ongoing research.

\section*{Acknowledgment}

The research presented in this paper was supported by the Danish Council for Independent Research (Det Frie Forskningsr\aa d) within the Sapere Aude Research Leader program, Grant No. 11-105159 ``Dependable Wireless Bits for Machine-to-Machine (M2M) Communications''.

\bibliographystyle{ieeetr}

\end{document}